\documentstyle[12pt]{article}

\begin{document}
\begin{titlepage}
\centerline {\bf Nuclear incompressibility: An analytical study on
leptodermous expansion}
\medskip
\centerline {V.S. Uma Maheswari, V.S. Ramamurthy and L. Satpathy}
\smallskip
\centerline {\it Institute of Physics, Bhubaneswar-751 005, India}
\smallskip
\begin{abstract}
{ A comparative study of the liquid-drop model (LDM) type expansions
of energy $E$ and compression modulus $K_A$ is made within the
energy density formalism using Skyrme interactions.
As compared to the energy expansion, it is found that,
in the pure bulk mode of density vibration, the LDM expansion of
$K_A$ shows an anomalous convergence behaviour due to {\it pair \
effect}. A least squares fit analysis is done to estimate the minimum
error, one would expect even with
synthetic data due to the inherent nature of the LDM expansion of
$K_A$ as well as the narrow range of accessible mass number $A$, in
the values of the various coefficients.
The dependence of the higher-order coefficients like
curvature and Gauss curvature on the coupling $\beta_c$ between the
bulk and surface parts of the monopole vibrations is analytically
studied. It is shown that the $K_A -$ expansion including the
dynamical effect ( $A-$ dependence of $\beta_c$ ) shows an `up-turn'
behaviour below mass number about 120, suggesting the inapplicability
of the LDM expansion of $K_A$ over this mass region.}

\end{abstract}
\end{titlepage}
\newpage

\centerline {\bf I.\ INTRODUCTION }
\medskip
Nuclear matter incompressibility $K_{\infty}$ is of fundamental
significance as it is an important ingredient in the nuclear equation
of state, which influences several astrophysical phenomena as well as
high energy heavy-ion collision processes. The nuclear breathing mode
or isoscalar giant monopole resonance (GMR) is the most promising
source of information on $K_{\infty}$.

	However, the nuclear breathing mode actually determines the
finite nuclear compression modulus $K_A$, and not $K_{\infty}$.
Hence, to determine the value of $K_{\infty}$ one has to extrapolate
from finite nuclei to nuclear matter, which is a highly non-trivial task.
Inspired by the success of the nuclear mass formulae, a liquid-drop
model (LDM) like approach is currently being adopted to extract
$K_{\infty}$ from its finite nuclear value $K_A$. In this approach,
one first expresses the finite nuclear compression modulus $K_A$ in
terms of the experimental breathing-mode energies $E_{gmr}$ using the
relation,
\begin{equation}
K_A = m<r^2> E_{gmr}^2/\hbar^2
\end{equation}
where the experimental values of the mean square radius $<r^2>$ are used;
and then one supposes an LDM expansion of $K_A$ to be valid, in
analogy to the semi-empirical mass formula;
\begin{equation}
K_A = K_v + K_s A^{-1/3} + K_c A^{-2/3} + K_{Cou} Z^2 A^{-4/3} +
K_{\beta}\beta^2
\end{equation}
where $\beta$ is the asymmetry parameter. The volume
coefficient $K_v$ determined from a fit to the available breathing
mode energies is identified with $K_{\infty}$, which is strictly
valid only in the scaling model.

Following this approach, Sharma {\it et al } \cite{Sh88,Sh89}
determined $K_{v}$
to be about 300 MeV using their data on various $Sn$ and $Sm$
isotopes, and on those of $^{24}Mg$, $^{90}Zr$ and $^{208}Pb$, which
was later contested by Pearson \cite{Pe91}.
Subsequently, Shlomo and Youngblood \cite{Sh93} made an extensive
least-squares fit
analysis by taking into account the available set of data on
breathing-mode numbering about 46, spread over a mass region $28
\le A \le 232$.
They concluded that this complete data set is not
adequate to determine the value of $K_{\infty}$ to better than a
factor of about 1.7  \ [ 200 -350 MeV ]. Further, they also observed in
their analysis that a free least-squares fit to Eq.(2) leads to errors
exceeding $100\%$ in all the coefficients.
As observed by Shlomo and Youngblood, such large errors may
arise partly due to the break-down of the scaling approximation over
light nuclei, deformation effects etc. Another reason may be that
the presently available data are inadequate for fixing all the
parameters in Eq.(2). We feel that, possible slow convergence and
correlations among the coefficients, will also contribute toward this
error.
However, it is not possible to pinpoint what fraction of the error
arises out of the quality and inadequacy of the data, and that which
arises due to the nature of convergence of the LDM expansion of $K_A$.
It is therefore desirable to make an extensive theoretical study of
the nature of convergence and the effect of correlations to arrive at
a conclusion.

Studies in this regard have been attempted[5-9] over the
years and have normally been done in the scaling model, since
an analytical relation between the experimental breathing-mode
energies and $K_A$ is available only within this model.
But, it is known from the works of Brack and his
collaborators[10-12] that there can exist several types of
density vibration, the scaling mode being a particular one.
It was also shown by them in a hydrodynamical approach that, the
degree of coupling between the bulk and surface parts of the density
vibrations,  sensitively depends upon the mass number $A$.

Let us consider the ground-state density of a nucleus to be given by
a Fermi-function ${\rho (r)}= \rho_{o}/[ {1+e^{{(r-R)\over a}}}]$
where $\rho_o$, $R$ and $a$ are the central density,
half density radius and diffuseness paramters respectively. When it
undergoes monopole vibrations, the density function will be
compressed or decompressed. In this state, it is assumed that the
density function can still be represented by a Fermi-function. Then,
following Brack and Stocker\cite{Br83}, one has
\begin{equation}
{\rho^c (r)}= {\rho_{c}\over {1+e^{{(r-R)\over \alpha_c}}}}
\end{equation}
where $\rho_c$ and $\alpha_c$ are defined as
\begin{eqnarray}
 \rho_{c} &=& \rho_{o} q \nonumber \\
 \alpha_c &=& a q^{\beta_c} =
a {\left ( {\rho_c / \rho_{o}} \right )}^{\beta_c}
\end{eqnarray}
The paramter $q$ represents the amount of compression or
decompression, $\beta_c$ gives the degree of coupling between the
bulk and surface parts of the density vibration and $\alpha_c$ is the
corresponding value of the surface diffuseness. The scaling property
of the half-density radius $R$ is defined by the number equation
given by Eqs.(6,12).  In literature[10-12], the mode of vibration
pertaining to $\beta_c =-1/3$ is normally referred to as scaling mode
of vibration.
In realistic hydrodynamical calculations \cite{Br83,Gl90}, which describe
the experimental data on breathing-mode reasonably well, it is found
that $\beta_c$ can vary from -0.23, in the case of $^{208}$Pb, to
-0.84 for $^{90}$Zr. Further, $\beta_c$ tends to $\pm
\infty$ for lighter nuclei like $^{48}$Ca  \ [$\beta_c$ = -24] and
$^{40}$Ca  \ [$\beta_c$ = +9.5], i.e. pure
surface like mode.

More importantly, when one makes use of an LDM expansion of
$K_A$  to fit the experimental GMR data, the various finite-size
coefficients in the expansion(2), such as surface $K_s$ and curvature
$K_c$, must be $A-$independent, as in the case of standard nuclear
mass formulae. Due to the  mass number dependence of the dynamical
coupling paramter $\beta_c$, there is still a residual mass
dependence in these coefficients obtained theoretically.
The asymptotic value of $K_s$ obtained in the limit $A
\longrightarrow \infty$ is about $-2 K_v$, and corresponds to the
pure bulk mode $\beta_c$=0.
This large, negative value of $K_s$ shall lead to a slower converging
$K_A -$ expansion as compared to the one obtained in the scaling
model, where $K_s \simeq - K_v$.
Therefore, it is of interest and relevance to study the convergence
properties of the LDM expansion of $K_A$ for this particular mode.
This has been stated\cite{Br83} to be a well
converging series, which may not be conclusive as we shall see.

In view of the above discussions, we make an analytical
study of the LDM expansion of $K_A$ derived with $\beta_c$=0. It is
found that this particular expansion shows
an anomalous convergence behaviour in the sense that certain
higher-order terms are equal in magnitude and opposite in sign to the
preceding lower-order terms in the
expansion, resulting in, what is termed here as, `pair effect'.
Secondly as the LDM expansion of $K_A$ is relatively new compared
to the well established energy expansion, it is worthwhile to make a
comparative study of both these expansions at a fundamental level.
Thirdly, we also make a least-squares fit analysis to get an idea
regarding the minimum error in the values of the various
coefficients, one would expect even with  synthetic data because of
the inherent nature of the LDM  expansion as well as, the inadequacy
of the narrow range of $A$ to fix the higher-order coefficients.
Finally, to ascertain the goodness of the $K_A-$ expansion for the
extraction of the various coefficients, we have studied the
the $\beta_c$ dependence of curvature $K_c$ and Gauss curvature $K_o$
coefficients, and thereby extending the pocket model calculations to
higher-order coefficients. It is found that, the validity of the LDM
expansion of $K_A$ in particular cases such as scaling and pure bulk
mode, does not necessarily quarantee unambiguous extraction of
$K_{\infty}$ in realistic situations.
This is shown by taking the dynamical effect (the $A-$
dependence of $\beta_c$) into account, where the nuclear
incompressibility $K_A$ is found to exhibit an `up-turn' behaviour
below mass number about 120, which may be of non-leptodermous origin.

\medskip
\centerline {\bf II.\ THE MODEL}
\medskip

Here, we analytically derive the LDM expansion of nuclear
incompressibility $K_A$ starting from first principles, and using the
concept of leptodermous expansion of energy. The model considered
here is essentially the same as the one given in Ref.\cite{Um92},
which we have now generalised to consider various modes of monopole
vibrations.

We consider a symmetric system of $A$ fermions, without the
Coulomb interaction, whose total energy $E$,
in the framework of energy density
formalism\cite{Ho64} is given by,
\begin{equation} E= \int \epsilon  \ [\rho (r)] d^{3}r \end{equation}
with the number constraint equation,
\begin{equation} A= \int \rho (r) d^{3}r  \end{equation}
where,
\begin{equation} \epsilon (r) = {{\hbar}^{2}\over 2m} \tau (r) + v(r)
\end{equation}

For the kinetic part $\tau (r)$, we use the extended Thomas-Fermi
functional\cite{Br85} upto $O(\hbar^{2})$, which for a symmetric system is
\begin{equation} \tau (r) = {3\over 5}{(1.5{\pi}^{2})}^{2/3}{\rho
}^{5/3}+{1\over 36}
{{(\nabla \rho )}^{2}\over \rho }+{1\over 3}{\nabla}^{2}\rho \end{equation}
For the potential part $v(r)$, we use the standard Skyrme forces,
without the spin-orbit contribution,
\begin{equation}
v(r)={3t_{0}\over 8}{\rho }^{2}+{t_{3}\over 16}{\rho }^{\sigma +2}+
{{3t_{1}+5t_{2}}\over 16}\tau \rho +{{9t_{1}-5t_{2}}\over 64}{(\nabla
\rho )}^{2} \end{equation}

Since our objective is to make an analytical study,
we do not attempt to solve the Euler-Lagrangian equations for
self-consistent densities. Further, to systematically study the
general nature of
the LDM compressibility coefficients and their dependence on the
coupling paramter $\beta_c$, it is desirable to parametrise the
density function $\rho (r)$ to be a Fermi function.
Therefore,  the ground-state density function $\rho (r)$
for a nucleus is given as
\begin{equation}
{\rho (r)}= {\rho_{o}\over {1+e^{{(r-R)\over a}}}}
\end{equation}
where $\rho_{o}$, $R$, and $a$ are the central
density, half-density
radius and the diffuseness parameter respectively.
In order to obtain analytical expressions for the various
finite-size compressibility coefficients
in terms of the coupling paramter $\beta_c$,
we define the general density distribution corresponding to a
compressed/decompressed state to be of the form given by Eq.(3).
Thus, our whole derivation is of general nature, and the ground-state
properties can be determined by imposing $q=\rho_c/\rho_o = 1$.

Normally, one expands
the finite nuclear ground-state energy into volume, surface etc using
Taylor's series. Then, each coefficient in the expansion is
calculated(for e.g., see  \cite{Na90})  making use of the semi-infinite
nuclear matter approximation . Instead, we here arrive
at the LDM expansion
of $K_A$ using the generalised Sommerfeld lemma \cite{Kr81,Sr82}.
One can then systematically express the total energy as a sum of volume,
surface, curvature and Gauss curvature.
Further, as we are interested in the convergence properties of
the LDM expansion of $K_A$, we go upto Gauss curvature order in both
the energy and compressibility expansions.

Then, the total energy $E$ at any state of compression/decompression
given by Eq.(5) results in the leptodermous expansion of $E$ as,
\begin{equation}
E = E_{v}{{4\pi R^{3}}\over 3}+E_{s}4\pi R^{2}+E_{c}8\pi
R+E_{o}4\pi \end{equation}
The terms $E_{v}$, $E_{s}$, $E_{c}$ and $E_{o}$ are the volume,
surface, curvature and Gauss curvature contributions to the total
energy $E$. Then, the half-density radius $R$ can be given in terms
of $\rho_{c}$, by using Eq.(6) as
\begin{equation}
R \simeq r_{oc}A^{1/3}{\left [ 1-{\pi^{2}\alpha_c^{2}\over
3r_{oc}^{2}A^{2/3}}
\right ] }  \end{equation}
where ${4\pi r_{oc}^{3}\over 3}{\rho_{c}} =1$. Subtituting the
above expression for $R$ in Eq.(11), and retaining terms upto the order of
$A^{-1}$, we obtain
\begin{equation}
{E\over A}=
e_{v}^{\ast}+e_{s}^{\ast}A^{-1/3}+e_{c}^{\ast}A^{-2/3}+e_{o}^{\ast}A^{-1}
\end{equation}
The expressions for the different coefficients in the above equation are
\begin{equation} e_{v}^{\ast} = p_{\alpha
}{\rho_{c}}^{2/3}+p_{0}{\rho_{c}}+p_{1}\alpha {\rho_{c}}^{5/3}+
p_{3}{\rho_{c}}^{\sigma +1} \end{equation}
\begin{eqnarray}
e_{s}^{\ast} &=& -(36\pi )^{1/3}\alpha_c (0.759p_{\alpha }\rho_{c}-
{p_{\beta}\over 2\alpha_c^{2}}{\rho_{c}}^{1/3}+p_{0}{\rho_{c}}^{4/3}+
1.359\alpha p_{1}{\rho_{c}}^{2} \nonumber \\
&\quad& -{ \ [p_2 + p_{1}(\beta -\gamma )]\over 6\alpha_c^{2}} {\rho_{c}}^{4/3}
-p_{3}g_{3s}^{\sigma }{\rho_{c}}^{\sigma +4/3})
\end{eqnarray}
\begin{eqnarray}
 e_{c}^{\ast} &=& 4\alpha_c^{2}(6{\pi}^{2})^{1/3}(
(1.517-{{\pi}^{2}\over 6})p_{\alpha }\rho_{c}^{4/3}+
{p_{\beta}\over 2\alpha_c^{2}}\rho_{c}^{2/3}+
(1.973-{{\pi}^{2}\over 6})p_{1}\alpha \rho_{c}^{7/3} \nonumber \\
&\qquad & +
(0.5g_{3c}^{\sigma }-{{\pi}^{2}\over 6})p_{3}\rho_{c}^{\sigma +5/3})
\end{eqnarray}
\begin{eqnarray}
e_{o}^{\ast} &=& -4\pi \alpha_c^{3}(
(2.602-{2{\pi}^{2}\over 3}0.759)p_{\alpha }\rho_{c}^{5/3}+
{{\pi }^{2}\over 6}{p_{\beta}\over \alpha_c^{2}}\rho_{c}+
(4.423-{2{\pi}^{2}\over 3}1.359)p_{1}\alpha \rho_{c}^{8/3} \nonumber \\
 &\quad & -
{\pi^{2}\over 3}p_{0}\rho_{c}^{2}
+{ \ [p_2+p_{1}(\beta- \gamma )]\over \alpha_c^{2}}({\pi^{2}\over 18}+{1\over
3})\rho_{c}^{2}
-(g_{3o}^{\sigma }-{2{\pi}^{2}\over 3}g_{3s}^{\sigma
})p_{3}\rho_{c}^{\sigma +2}  )
\end{eqnarray}
where $p_{\alpha }={{\hbar^2\over 2m} }\alpha$,$p_{\beta
}={{\hbar^2\over 2m}} \beta $,
$p_{0}={3t_{0}\over 8}$, $p_{3}={t_{3}\over 16}$,
$p_{1}={3t_{1}+5t_{2}\over 16}$,
$p_{2}={9t_{1}-5t_{2}\over 64}$, and $\alpha ={3\over
5}(1.5\pi^{2})^{2/3}$, $\beta={1\over 36}$ and $\gamma= {1\over 3}$.
The quantities $g_{3s}^{\sigma }$, $g_{3c}^{\sigma }$ and
$g_{3o}^{\sigma }$ are calculated using the integral
\begin{equation}
\eta_{\nu }^{(k)}=(-1)^{k}{\int_{0}^{\infty }u^{k}\left [
{1+(-1)^{k}e^{-u\nu }\over (1+e^{-u})^{\nu }}-1 \right ] du }
\end{equation}
where $\nu =\sigma +2$, and $g_{3s}^{\sigma }=\eta_{\nu}^{(0)}$,
$g_{3c}^{\sigma }=2\eta_{\nu}^{(1)}$ and  $g_{3o}^{\sigma }=\eta_{\nu}^{(2)}$.

 It must be noted
that all the coefficients in Eq.(13) are function of the
ground-state values of
central density $\rho_{o}$ and the diffuseness parameter $a$, and
also are function of the
compression variable $q$ and the coupling paramter $\beta_c$.
$\rho_o$ and $a$ are normally determined using the energy
minimisation criteria.
Due to the finite compressibility of the nuclear fluid, both the
quantities could be functions of the size of the system, characterised by their
mass number $A$. Therefore, Eq.(13) can not be identified as the LDM
expansion of the nuclear energies.

To extract the $A$-dependence of the energy coefficients in
Eq.(13), we define $\rho_o (A) = \rho_{\infty} + \delta \rho$, where
$\rho_{\infty}$ is the symmetric infinite nuclear matter saturation density.
$\rho_{o}(A)$ can be determined using the saturation condition,
\begin{equation}
{d{(E/A)}\over d\rho }\mid_{\rho_{o}}=0
\end{equation}
Remembering that the above derivative is a total one, we define
\begin{equation}
\rho_o \ [{d{(E/A)}\over d\rho_c }]_{\mid_{\rho_c=\rho_{o}}}
= [{d{(E/A)}\over dq }]_{\mid_{q=1} }
\end{equation}
The saturation condition given by Eq.(19) now becomes,
\begin{equation}
e_{v}^{\prime }+e_{s}^{\prime }A^{-1/3}+e_{c}^{\prime
}A^{-2/3}+e_{o}^{\prime }A^{-1}
 + e_{f}^{\prime }A^{-4/3} + e_{h}^{\prime }A^{-5/3} = 0
 \end{equation}
Here, the prime denotes total differentiation of $e_{i}^{\ast}$'s
with respect to  the
central density variable. To determine all the coefficients at the
density
$\rho_{\infty}$ of symmetric infinite nuclear matter , we use the definition
\begin{equation}
\rho_{o}=\rho_{\infty}+\delta \rho
\end{equation}
where $\delta \rho $ contains all the possible finite-size effects.

Making Taylor's expansion of all the coefficients in Eq.(21) around
$\rho_{\infty}$ upto $O(\delta \rho)^2$, we obtain,
\begin{equation}
\delta \rho = -{\left [  {
e_{s}^{\prime }A^{-1/3}+
e_{c}^{\prime }A^{-2/3}+
e_{o}^{\prime }A^{-1}
\over e_{v}^{\prime \prime }+
e_{s}^{\prime \prime }A^{-1/3}+
e_{c}^{\prime \prime }A^{-2/3}+
e_{o}^{\prime \prime }A^{-1}+
0.5(e_{v}^{\prime\prime \prime }+e_{s}^{\prime \prime\prime
}A^{-1/3}+ \cdots )\delta \rho
} \right ] }_{ \mid_{\rho_{\infty}}}
\end{equation}
As the above equation contains $\delta\rho$ on the right hand side
also, we determined the same by an iterative method. The zeroth order
expression of $\delta\rho$ is equal to the above equation without the
bracketed term in the denominator.

To arrive at the LDM expansion of $E/A$ and $K_A$, we should be
able to expand $\delta \rho $ in powers of $A^{-1/3}$.
Unlike in Ref. \cite{Um92}, here we consider the influence
of higher-order terms like $e_{s}^{\prime\prime}$,
$e_{c}^{\prime\prime}$ {\it etc} by making a binomial expansion of
the denominator in Eq.(23).
Considering terms upto third order, we have,
\begin{eqnarray}
 &\qquad& {\left [  { 1+
({e_{s}^{\prime\prime}\over e_{v}^{\prime\prime}}-
{e_{v}^{\prime\prime\prime}e_s^{\prime}\over
2(e_{v}^{\prime\prime})^2}) A^{-1/3}+ \cdots
 } \right ]}^{-n} \nonumber \\
 &\qquad& \qquad
  \simeq {1+f_{1}(n)A^{-1/3} +f_{2}(n)A^{-2/3}+ \cdots}
\end{eqnarray}
where,
\begin{eqnarray}
f_{1}(n) &=& -n \left [  {e_{s}^{\prime\prime}\over e_{v}^{\prime\prime}}
-{e_{v}^{\prime\prime\prime}e_s^{\prime}\over
2(e_{v}^{\prime\prime})^2} \right ]
\nonumber \\
f_{2}(n) &=& -n \left [  {e_{c}^{\prime\prime}\over e_{v}^{\prime\prime}}
-{e_{s}^{\prime\prime\prime}e_s^{\prime}\over
2(e_{v}^{\prime\prime})^2}
-{e_{v}^{\prime\prime\prime}e_c^{\prime}\over
2(e_{v}^{\prime\prime})^2} \right ] \nonumber \\
 &\quad & -n \left [
{e_{v}^{\prime\prime\prime}e_s^{\prime\prime}e_s^{\prime}\over
2(e_{v}^{\prime\prime})^3}
-{e_{v}^{\prime\prime\prime\prime}{e_s^{\prime}}^2\over
6(e_{v}^{\prime\prime})^3}
 \right ] \nonumber \\
 &\quad&  + {n(n+1)\over 2} {\left [  {e_{s}^{\prime\prime}\over
e_{v}^{\prime\prime}}
-{e_{v}^{\prime\prime\prime}e_s^{\prime}\over
2(e_{v}^{\prime\prime})^2}
\right ] }^2 \nonumber
\end{eqnarray}
It may be recalled here that prime denotes total differentiation with
respect to the central density variable for a fixed value of
$\beta_c$. One then has for any arbitrary function
$f \ [\rho_c, \alpha_c (\rho_c )]$,
\begin{eqnarray}
f^{\prime} &\equiv & \left ( {df\over d\rho_c }\right ) \mid_{\rho_o} =
{1\over \rho_o}\left ( {df\over dq }\right ) \mid_{q=1}
\nonumber \\
f^{\prime\prime} &\equiv & \left ({d^2f\over d\rho_c^2 }\right )
\mid_{\rho_o} =
{1\over \rho_o^2}\left ( {d^2f\over dq^2 }\right ) \mid_{q=1}
\end{eqnarray}

\vskip 0.5 true cm
\centerline {\bf A. LDM expansion of energy }
\medskip

Now, the LDM expansion of $E/A$ can be obtained in a
straight-forward manner by performing Taylor's expansion of each of
the term in Eq.(13) around $\rho_{\infty}$ as,
\begin{eqnarray}
{E\over A} &=&
e_{v}^{\ast}(\rho_{\infty})+e_{s}^{\ast}(\rho_{\infty})A^{-1/3}+
\cdots + e_{o}^{\ast}(\rho_{\infty})A^{-1} \nonumber \\
&\quad & + \left (
{e_{s}}^{\prime}(\rho_{\infty})A^{-1/3}+ \cdots
+{e_{o}}^{\prime}(\rho_{\infty})A^{-1} \right )\delta\rho \nonumber
\\ &\quad & + {1\over 2}\left ( {e_{v}}^{\prime \prime}(\rho_{\infty})+
{e_{s}}^{\prime\prime }(\rho_{\infty})A^{-1/3}+ \cdots
+{e_{o}}^{\prime\prime }(\rho_{\infty})A^{-1} \right )
{\left(\delta\rho \right )}^2 \nonumber \\
&\quad & + {1\over 6}\left ( {e_{v}}^{\prime\prime\prime}(\rho_{\infty})+
{e_{s}}^{\prime\prime\prime }(\rho_{\infty})A^{-1/3}+ \cdots
+{e_{o}}^{\prime\prime\prime }(\rho_{\infty})A^{-1} \right )
{\left(\delta\rho \right )}^3 + \cdots
\end{eqnarray}
Then, by using Eqs.(23-24) in the equation (26), and
grouping terms with same power of $A$, the complete LDM expansion of
$E/A$ upto $O(A^{-1})$ is
\begin{equation}
{E\over A} = a_{v}+a_{s}A^{-1/3}+a_{c}A^{-2/3}+a_{0}A^{-1}
\end{equation}
where the various LDM coefficients in the above equation are defined
as,
\begin{eqnarray}
 a_{v} &=& e_{v}^{\ast}(\rho_{\infty}) \nonumber \\
 a_{s} &=& e_{s}^{\ast}(\rho_{\infty}) \nonumber \\
 a_{c} &=& e_{c}^{\ast}(\rho_{\infty})+
           \left (d_{11}e_{s}^{\prime}\right )+
          {1\over 2}\left (d_{22}e_{v}^{\prime\prime}\right )
\nonumber \\
 a_{0} &=& e_{0}^{\ast}(\rho_{\infty})+
\left ( d_{11}e_{c}^{\prime}+d_{12}e_{s}^{\prime} \right ) \nonumber \\
&\quad & +
{1\over 2}\left (d_{22}e_{s}^{\prime\prime}+d_{23}e_{v}^{\prime\prime}\right)
+{1\over 6}\left (d_{33}e_{v}^{\prime\prime\prime}\right )
\end{eqnarray}
It may be noted that unlike the leptodermous expansion given by
equation (11), the LDM expansion of energy is an infinite series.
This infinite nature comes due to the $A^{-1/3}$ expansion of $\delta
\rho$, which is derived by making a binomial series. Therefore, we
feel it is essential
to go upto atleast Gauss curvature order in the LDM expansions of
energy and compressibility, while one studies their convergence
properties.

The explicit expressions for the various factors $d_{11}$,$d_{12}$
{\it etc} are;
\begin{eqnarray}
(i)  \quad d_{11} &=& -{1\over e_{v}^{\prime\prime}}e_{s}^{\prime}
\nonumber \\
d_{12} &=& -{1\over e_{v}^{\prime\prime}}
       \left ( e_{c}^{\prime}+e_{s}^{\prime}f_{1}(1)\right )
 \nonumber \\
 d_{13} &=& -{1\over e_{v}^{\prime\prime}}
 \left ( e_{0}^{\prime}+e_{c}^{\prime}f_{1}(1)+e_{s}^{\prime}f_{2}(1)\right )
\nonumber \\
(ii)  \quad d_{22} &=& {1\over {e_{v}^{\prime\prime}}^2}{e_{s}^{\prime}}^2
\nonumber \\
d_{23} &=& {1\over {e_{v}^{\prime\prime}}^2}\left (
2e_{s}^{\prime}e_{c}^{\prime} +
{e_{s}^{\prime}}^2f_{1}(2) \right )
\nonumber \\
(iii)  \quad d_{33} &=& -{1\over {e_{v}^{\prime\prime}}^3}
{e_{s}^{\prime}}^3 .
\end{eqnarray}
The important point to be noted regarding
these factors is that they all have an explicit dependence upon
$K_{\infty}$, which in turn controls the convergence behaviour of the
LDM expansion. However, in the case of energy, the two leading
coefficients volume and surface are independent of these factors, and
thereby $K_{\infty}$. Only the higher-order terms like curvature and
Gauss curvature are dependent on $K_{\infty}$.

\vskip 0.5 true cm
\centerline {\bf B. LDM expansion of nuclear compression modulus $K_A$}
\medskip

In the following paragraphs, we derive the LDM expansion of $K_A$ .
The finite nuclear compression modulus $K_A$ is calculated using the
definition,
\begin{equation}
K_{A}=9\rho_{o}^{2}{\left ( {d^{2}{E/A}\over d\rho_c^{2}}\right ) }_
{\mid_{\rho_c = \rho_{o}} }
 = 9 {d^{2}{E/A}\over dq^{2}}\mid_{q=1}
\end{equation}
for a given value of $\beta_c$.

Making use of equations (27-28) in the above definition, we have
\begin{equation}
K_{A}=K_{v}^{\ast}(\rho_{o})+K_{s}^{\ast}(\rho_{o})A^{-1/3}+
K_{c}^{\ast}(\rho_{o})A^{-2/3}+K_{o}^{\ast}(\rho_{o})A^{-1}
\end{equation}
where,
\begin{eqnarray}
K_{i}^{\ast}(\rho _{o}) &=& 9\rho_{o}^{2}{d^{2}e_{i}^{\ast}\over
d\rho_c^{2}}\mid_{\rho_{o}} \nonumber \\
\quad &=&  9{d^{2}e_{i}^{\ast}\over dq^{2}}
\mid_{q=1} \qquad ; \quad i=v,s,c,\quad {\rm and} \quad o
\end{eqnarray}
Now, the LDM expansion $K_{A}$ can be obtained in a
straight- forward manner by performing Taylor's expansion of each of
the term in Eq.(31) around $\rho_{\infty}$ as,
\begin{eqnarray}
K_{A} &=&
K_{v}^{\ast}(\rho_{\infty})+K_{s}^{\ast}(\rho_{\infty})A^{-1/3}+
\cdots + K_{o}^{\ast}(\rho_{\infty})A^{-1} \nonumber \\
&\quad & + \left ( {K_{v}^{\ast}}^{\prime}(\rho_{\infty})+
{K_{s}^{\ast}}^{\prime}(\rho_{\infty})A^{-1/3}+ \cdots
+{K_{o}^{\ast}}^{\prime}(\rho_{\infty})A^{-1} \right )\delta\rho
\nonumber \\
&\quad & + {1\over 2}\left ( {K_{v}^{\ast}}^{\prime \prime}(\rho_{\infty})+
{K_{s}^{\ast}}^{\prime\prime }(\rho_{\infty})A^{-1/3}+ \cdots
+{K_{o}^{\ast}}^{\prime\prime }(\rho_{\infty})A^{-1} \right )
{\left(\delta\rho \right )}^2 \nonumber \\
&\quad & + {1\over 6}\left (
{K_{v}^{\ast}}^{\prime\prime\prime}(\rho_{\infty})+
{K_{s}^{\ast}}^{\prime\prime\prime }(\rho_{\infty})A^{-1/3}+ \cdots
+{K_{o}^{\ast}}^{\prime\prime\prime }(\rho_{\infty})A^{-1} \right )
{\left(\delta\rho \right )}^3 + \cdots
\end{eqnarray}
Here, we have included all the terms upto $O{(\delta\rho)}^3$, so
that correct estimates upto Gauss curvature order {\it i.e.}
$O(A^{-1})$, can be obtained, which is desirable in this study of the
convergence behaviour of the LDM expansion of $K_{A}$.

Then, by using Eqs.(23-24) in the equation (33) and
grouping terms with same power of $A$, the complete LDM expansion of
$K_{A}$ upto $O(A^{-1})$ is
\begin{equation}
K_{A} = K_{v}+K_{s}A^{-1/3}+K_{c}A^{-2/3}+K_{o}A^{-1}
\end{equation}
where the various LDM coefficients in the above equation are defined
as,
\begin{eqnarray}
 K_{v} &=& K_{v}^{\ast}(\rho_{\infty}) \nonumber \\
 K_{s} &=& K_{s}^{\ast}(\rho_{\infty})+
 \left (d_{11}K_{v}^{\prime}\right ) \nonumber \\
 K_{c} &=& K_{c}^{\ast}(\rho_{\infty})+
           \left (d_{11}K_{s}^{\prime}+d_{12}K_{v}^{\prime}\right )+
          {1\over 2}\left (d_{22}K_{v}^{\prime\prime}\right
)\nonumber \\
 K_{o} &=& K_{o}^{\ast}(\rho_{\infty})+
\left (
d_{11}K_{c}^{\prime}+d_{12}K_{s}^{\prime}+d_{13}K_{v}^{\prime} \right
) \nonumber \\ &\quad & +
{1\over 2}\left (
d_{22}K_{s}^{\prime\prime}+d_{23}K_{v}^{\prime\prime}\right )
+{1\over 6}\left ( d_{33}K_{v}^{\prime\prime\prime}\right )
\end{eqnarray}
The explicit expressions for the various factors $d_{11}$,$d_{12}$
etc are as given in Eq.(29). It may be noted that unlike in
the case of energy expansion, all the finite-size coefficients ($K_s$,
$K_c$, $K_o$) explicitly depend upon $K_{\infty}$ through the factors
$d_{11}$, $d_{12}$ etc.

\medskip
\centerline {\bf {III. RESULTS AND DISCUSSIONS }}
\medskip

In this section, we discuss the basic nature of an LDM expansion of
energy and incompressibility. In addition, the dependence of the
various compressibility LDM coefficients on the coupling parameter
$\beta_c$ is also studied.

\medskip
\centerline {\bf A. Structure of the LDM coefficients and dependence
on $\beta_c$ }
\medskip

The first and the important point is that in the case of
energy, the bulk part totally decouples itself from all the
surface effects due to the saturation
condition $e_{v}^{\prime}(\rho_{\infty}) = 0$, in the leading
order $\delta\rho$, which is evident from
Eq.(26). This, in turn, results in the two important LDM energy
coefficients $a_{v}$ and $a_{s}$, given in Eq.(28), being pure
in nature, in the sense that there are no contributions arising from
$\delta\rho$. In contrast, in the case of compressibility, the volume
gets strongly coupled to all the surface effects even through the
leading order in $\delta\rho$, and hence, the surface coefficient
$K_{s}$ has an additional term $d_{11}K_{v}^{\prime}$, which
explicitly depends upon $K_{\infty}$ through $d_{11}$.
This fact contributes to the essential difference between the LDM
expansion of $E/A$ and $K_{A}$. As it can be seen from Eqs.(28) and
(35), it also gives rise to additional terms in the higher-order
coefficients in the expansion of $K_{A}$. With this observation, we
now calculate the values of the various LDM coefficients in Eqs.(27)
and (34).

We calculated the various LDM energy coefficients using the
analytical expressions given in Eq.(28) for the three Skyrme forces
SkM$^*$, SkA and S3. In this calculation, only the curvature $a_c$
and Gauss curvature $a_o$ coefficients have a $\beta_c$ dependence.
We found that both these coefficients are almost invariant with
respect to $\beta_c$. Therefore, we have used $\beta_c =0$ while
calculating the energy coefficients, as is normally done.
Values obtained for the LDM energy coefficients using Eq.(28) are
given in Table I for the three Skyrme forces.
The values obtained here for the various LDM coefficients agree
qualitatively well with those found in literature \cite{Br85}.
In regard to the nuclear curvature, it may be noted that
the semi-classical ETF models \cite{St88,Du91} give a value of
about 10 MeV, as against the
value of about zero, determined \cite{Mo88} from experiments. This is
the so-called
nuclear curvature energy anomaly, which is yet to be resolved.

Similarly, we calculate the values of compressibility coefficients
using Eq.(35). In this calculation, we have shown our results for
three particular values of $\beta_c$ in Table II.
The ratio $\mid K_s/K_v \mid $ is also tabulated. It can be seen that
as $\beta_c$ decreases from
$\beta_c =0$ to $\beta_c = -1/2$, $\mid K_s/K_v \mid $ also decreses.
This behaviour of $K_s$ is in agreement with that of the pocket model
\cite{Br82}. Further, we find the scaling model result $K_s
\simeq - K_v$
is found to be well satisfied for $\beta_c = -1/3$.

In our study we have also calculated the higher-order coefficients
like $K_c$ and $K_o$ for different values of $\beta_c$.
The value of $K_c$ remains more or less unchanged with respect to
$\beta_c$. Due to this, as $\beta_c$ becomes more and more negative,
curvature effect becomes as important as the surface. This may be due
to the fact as $\beta_c$ becomes more and more negative, the
finite-size effects play a progressively important role in the
breathing vibrations of a nucleus.
Further, from the values of all finite-size
compressibility coefficients like $K_s$, $K_c$ etc given in Table
II, it can be seen
that $\beta_c \ge 0 $ will lead to a relatively slow converging
series compared to the other negative values of $\beta_c$. In other
words, as $\beta_c$ becomes more and more positive, $\mid K_s / K_v
\mid $ increases.

It may be mentioned here that in Ref. \cite{Um92}, we
had calculated the coefficients using $\beta_c=0$ and neglecting
terms like $e_s^{\prime\prime}A^{-1/3}$ ,
$e_c^{\prime\prime}A^{-2/3}$ etc in the denominator of Eq.(23). Now
retaining those terms and performing binomial expansion as given in
Eq.(24), we find that the values of $K_c$ and $K_o$ obtained with
$\beta_c$=0 get substantially modified.
For reasons discussed in the introduction, we presently study the
convergence properties of the
leptodermous expansion of $K_A$ with $\beta_c =0$, and also make a
comparative study of the LDM expansions of energy and compressibility.

\medskip
\centerline {\bf B. Convergence behaviour and Pair effect }
\medskip

To study the convergence of the LDM expansion of both energy and
compressibility, it is necessary to evaluate exactly the values of
energy $E$ and compressibility $K_{A}$ corresponding to
symmetric(N=Z) systems, which will then be compared
with the sum of all the terms given by the right hand side of
Eqs.(27) and (34) respectively. We determine the total energy
numerically
using Eqs.(5-10) and then, $K_{A}$ can be obtained using
Eq.(30) with $\beta_c = 0$.
Here, the diffuseness parameter $a$ is held fixed at its semi-infinite
nuclear matter(SINM) value.
The values so obtained for four representative nuclei in the range
$40 \le A \le 200 $ are presented in Tables III and IV
for the energy and compressibility respectively for Skyrme forces
SkM$^*$,SkA and S3.
It can be seen that in the case of energy, both the results agree
extremely well,{\it i.e.} upto about second decimal place, and even,
in the case of compressibility the agreement is quite good. This
result implies that the convergence of the LDM expansion of $K_{A}$
is almost as good as the LDM expansion of $E/A$.

To understand this inference in a better way, we have presented in
Tables V and VI,
the contributions of the successive terms in the energy and
compressibility expansions respectively, for the three forces
and, for two representative mass numbers A=40 and 200. It can be
seen that in the case of energy, the successive terms decrease by a
factor of about 5, giving rise to a rapidly converging series . In
contrast to this, we find in the case of compressibility, \ [see
Table VI ], the Gauss curvature term $K_{0}A^{-1}$ is of the same
order as the curvature term $K_{c}A^{-2/3}$. In fact, for values
of $A$ below about 150 ,$K_{0}A^{-1}$ even overshoots
$K_{c}A^{-2/3}$. Same is also found to be true in our calculation for
the higher-order terms $K_{h}A^{-5/3}$ and $ K_{f}A^{-4/3}$.
Further, it is interesting to note that
$K_{c}A^{-2/3}$ and $K_{0}A^{-1}$ are almost equal in magnitude, but
opposite in sign, which is also the case with  $K_{f}$ and
$K_{h}$. Hence, we find that
the terms of higher order than the surface one in the LDM expansion
of $K_A$ cancel pairwise.
 This {\it pair} effect gives rise to a misleading conclusion
regarding the importance of higher-order terms in the LDM expansion
of $K_{A}$, unless otherwise investigated. Thus, the LDM expansion of
$K_{A}$ in the case of pure bulk mode shows an anomalous behaviour,
in the sense that $K_{0}A^{-1}
\simeq -K_{c}A^{-2/3}$, in contrast to the rapidly converging energy
expansion.

In addition, it may be mentioned here that the convergence properties
of the LDM expansion of $K_A$ in the case of scaling
mode has been widely studied[6-9], going upto order of curvature
term. In our study, including terms upto Gauss curvature order, we
arrive at similar conclusions. Although, {\it pair} effect is not
observed in the scaling case, the rate of convergence is still found
to be relatively slow as compared to the energy expansion.

Further, we would like to state that the above results
were found to remain valid when one uses more realistic ETF
functionals including $\hbar^4$ terms, and with a generalised
Fermi-function for $\rho (r)$.
In view of this new feature in the convergence behaviour of the LDM
expansion of $K_A$, the question arise:
whether we can extract the various
coefficients in the LDM expansion of $K_A$ using a least squares fit analysis,
which may manifest the effect of correlations amongst the coefficients.

\medskip
\centerline {\bf IV. LEAST SQUARES FIT ANALYSIS }
\medskip

In the earlier section, we have obtained the exact values of energy
$E/A$ and compressibility $K_{A}$ numerically and also,
the values for the various LDM
coefficients in their expansions for SkM$^*$,SkA and
S3 forces, and have established their goodness.
Now, in this section, we would like to determine these
coefficients again from a least-squares fit to the
exact values of $E/A$ and $K_{A}$ (referred to as
synthetic data) for a set of nuclei. A
comparison of the two sets of values will then demonstrate clearly
the reliability of the LDM expansions of $E/A$ and $K_{A}$ for the
extraction of nuclear matter properties and, the surface properties of
finite nuclei. This will also establish the convergence behaviour of
both the expansions.

\vskip 0.5 true cm
\centerline {\bf A. For symmetric case}
\medskip

Here, we attempt to determine all the coefficients in the LDM
expansion of $K_A$ pertaining to  symmteric systems(N=Z) by making a free
least squares fit to the numerically calculated values of $K_A$ using
Eqs.(5-10,30), for 210 nuclei in the mass range $40 \le A \le 250 $.
In Tables VII and VIII, we have presented the results so obtained for
$E/A$ and $K_{A}$ respectively, for the SkA force only,
whose value of $K_{\infty}$ lies in between those of SkM$^*$ and S3 forces.
The corresponding error in all the coefficients calculated using the
standard method \cite{Be69} are also given.
The first row shows the exact values of the coefficients obtained in
the previous section, which are presented for comparison.

It can be seen from Tables VII and VIII that, in the case of both
energy and compressibility, the volume term is quite well determined
and its value progressively improves with the inclusion of higher-order
terms. Same is almost true for the surface coefficients. And, the
higher-order terms such as curvature and Gauss curvature are
relatively ill-determined. This is presumably due to correlations
amongst the various coefficients, which is clear from Eqs.(28) and (35).
Thus, we find that in the case of symmetric systems without the
Coulomb effect, the principal coefficients like volume
compressibility $K_v$ and surface compressibility $K_s$ can be
extracted reliabily by means of a least square fit.

\vskip 0.5 true cm
\centerline {\bf  B. Inclusion of Coulomb and asymmetry effects }
\medskip

The Coulomb force plays an important role in real nuclei, and
also, because of its influence in the extraction of $K_{\infty}$ from
the breathing-mode data as shown by Pearson \cite{Pe91} and Shlomo $\&$
Youngblood \cite{Sh93}, it is worthwhile to consider the effect of
Coulomb interaction in our present analysis.

To investigate this, we repeat our calculation
including the Coulomb force and asymmetry effect, for many realistic nuclei
lying within the mass region $40\le A \le 250$.
In this calculation of total energy(5), we suppose the neutron and
proton density distributions at the ground-state to be of the form,
\begin{equation}
{\rho_l (r)}= {\rho_{ol}\over {1+e^{{(r-R_l)\over a_l}}}} \quad ;
\quad l=n,p
\end{equation}
where the various parameters are as defined in Eq.(10), and are
determined by energy minimisation criteria.
Also, for the sake of simplicity, we have used the same value of
diffuseness parameter for both proton
and neutron density distributions, $a_n=a_p$. The Coulomb energy is
given as $E_{Cou} = 0.6Z^2e^2/R$, where we have considered the
nucleus to be a uniformly charged sphere of radius $R$ as is
normally done\cite{Bl81,Na90}.
One can then calculate $K_A$ using the total
energy expression(5) including asymmetry and Coulomb effects in Eq.(30).

Now, we make a least squares fit to the so-obtained theoretical
values of $K_A$ numbering 210.
The results so obtained from the fit for, the case of
compressibility is given in Table IX for the SkA force.
We find that, in the 4-paramter fit involving $K_v$, $K_s$, $K_{Cou}$
and $K_{\beta}$, the extraction of all these coefficients, except for
the symmetry compressibility $K_{\beta}$, are reliable.
The discrepancies in $K_v$ and $K_s$ as compared to the exact values
are about $4\%$ and $12\%$ respectively.
With the introduction of a surface-asymmetry $K_{s\beta}$ term,
i.e. in the 5-paramter fit, the estimate for $K_{\beta}$ improves,
however, at the cost of important coefficients like $K_v$ , $K_s$ and
$K_{Cou}$. To see the effect of a curvature term, we made a
5-parameter fit ($K_v$, $K_s$, $K_{Cou}$,$K_{\beta}$ $\&$ $K_c$) to
the 210 model data on  $K_A$. As it can be seen,
introduction of a curvature term as a free parameter somewhat spoils
the whole fit,
introducing a maximum of discrepancy of about $40\%$ in $K_v$.

Therefore, from our analysis using synthetic data, it can be concluded
that inspite of the correlations among various coefficients, it may be
possible to extract $K_{\infty}$. However, it must be mentioned that
even with such synthetic data, inclusion of higher-order terms like
$K_c$ and $K_{s\beta}$ as free parameters in the fit, lead to poorer
determination of the volume coefficient $K_{\infty}$, and also other
coefficients. Thus, the present theoretical study clearly points out
the goodness and limitations of the LDM expansion of $K_A$ and the
extent of error inherently present in this approach, in regard to the
determination of $K_{\infty}$.

\medskip
\centerline {\bf V. DYNAMICAL EFFECTS }
\medskip

Normally, the convergence properties of the LDM expansion of $K_A$
are mostly studied within the scaling model. This is because, only
within this model, one can easily relate $K_A$ to the experimental
data on GMR. However, in general, the mode of density vibrations is
neither scaling-like nor pure bulk-like. Hence, we need to examine
the convergence properties of $K_A-$ expansion taking into account
this dynamical effect.

To do so, firstly, we need to relate to the GMR data for any mode of
monopole vibrations. In other words, we need to find a general
empirical relation analogous to Eq.(1), and thereby, one
can obtain experimental $K_A$ values from the GMR data without making
the scaling assumption. Once $K_A$ values are known from the GMR
data, one may then use Eq.(2) to extract $K_{\infty}$. Hence, in the
following, we address two aspects: Firstly, how can one obtain a
general relation between $K_A$ and experimental $E_{gmr}$ ?,
Secondly, how does $K_A$ behave under the most general conditions?

In the appendix, we discuss the above mentioned first aspect using the
hydrodynamical approach. This justifies our study of the $K_A-$
expansion in the generalised situation, i.e. without using the
scaling assumption. With this, we now focus upon the second aspect,
i.e. the general behaviour of $K_A$ taking into account the $A-$
dependence of $\beta_c$, in the following.

Within our analytical model, we calculate realistic values of $K_A$
using the general expression
\begin{equation}
K_A (\beta_c ) = K_v + K_s ( \beta_c ) A^{-1/3} + K_c ( \beta_c )
A^{-2/3} +K_{Cou}Z^2A^{-4/3} + K_{\beta} \beta^2
\end{equation}
where $K_{Cou}$ $\&$ $K_{\beta}$ are respectively the Coulomb and
asymmetry compressibility coefficients and the asymmtery paramter
$\beta = (N-Z)/A $. The values for $K_{Cou}$ and $K_{sym}$ obtained
with SkM$^*$ force are $-$4.70 MeV and $-$349.0 MeV. In the calculation of
$K_{Cou}$, we have considered only the direct Coulomb term $a_C =
0.6e^2/r_o$. Realistic values of $K_s$ and $K_c$ are obtained as
follows.

For a given nucleus (A,Z), the optimum value of
$\beta_c$ that will give rise to an excitation energy $\hbar \omega$
close to the experimental value can
be obtained using\cite{Br83},
\begin {equation}
\beta_c = {\bar \beta} - \sqrt {{\bar \beta}^2 + 1}
\end{equation}
where ${\bar \beta}$ is related to $A$ as
\begin{equation}
{\bar \beta} = 0.685 A^{1/3} - 2.15
\end{equation}

So, for a given A and the
corresponding $\beta_c$ obtained from Eq.(37), we have calculated
$K_s$ and $K_c$ for the SkM$^*$ force using Eq.(38).
The values thus obtained for $K_s$
and $K_c$ are plotted as a function of $\beta_c$ in Figs.(1) and
(2) respectively.
It can be seen that as $\beta_c$ varies from $-$0.6 to 0.0, $K_s/K_v$
varies from from about $-$0.5 to $-$2. On the other hand, variation of
$K_c$ is not so prominent. Further, it is interesting to note that
value of $K_c $ shows a minimum at about the scaling model
value ($\beta_c$ = $-$1/3).
An important point to be noted is that since $\beta_c$ is A-dependent,
and $K_s$ $\&$ $K_c$ are $\beta_c-$dependent, the surface and
curvature compressibility coefficients have a residual mass
dependence.
To obtain the unique A$-$independent value of $K_s$ and $K_c$, one
should take the limit $A \longrightarrow \infty$, which leads to
$\beta_c =0$.
The asymptotic value of $K_s$ thus obtained is nothing but the value
$K_s \simeq -2K_v $, as noted earlier\cite{Sh89}.
And, the true asymptotic value of curvature compressibility
coefficient is the value of $K_c$ obtained with $\beta_c = 0$ plus the
contribution coming from the $\beta_c$ dependence of $K_s$.

Using the values of $K_s$ and $K_c$ calculated
for the SkM$^*$ force with the optimum values of $\beta_c$ in
Eq.(38), we plot in Fig.(3), the so-determined
$K_A ( \beta_c )/K_v $ values as a function of $A^{-1/3}$. For sake of
comparison, we have shown in Fig.(4) the values of $K_A (\beta_c)/K_v$
obtained for three particular values of $\beta_c$. It may be
mentioned that, for $A \ge 250$, the Coulomb force is switched off and N=Z.
It can be clearly seen from Fig.(3) that $K_A$
shows an ` up - turn ' behaviour as against the nice linear behaviour
found for the three particular values of $\beta_c$ in Fig.(4). This up-turn
behaviour or change in the slope may be suggesting the onset of the
breakdown of the leptodermous expansion of $K_A$ below mass number
approximately 120.
Indications for such an increasing nature of $K_A$ can also be found from
the hydrodynamical calculations \cite{Br83}.

In analysing real data, one should indeed expect such an `up-turn'
behaviour as against the nice linear behaviour obtained under scaling
or pure bulk mode assumption, which suggests that higher-order
coefficients like $K_c$ become important over medium and low mass regions.
Because of this `up-turn' behaviour , it may be difficult to
consistently determine all the parameters in Eq.(2) from a
fit to the presently available few tens of data on
GMR, which in-turn shall impair the extraction of the important
quantity, $K_{\infty}$.

\medskip
\centerline {\bf VI. CONCLUSION}
\medskip
In conclusion, the LDM expansion of $K_A$ in the pure bulk mode,
shows an anomalous convergence behaviour due to pair effect, as
compared to the widely studied scaling mode. It is also found that
the $K_A-$ expansion in both the cases is relatively quite slow, as
compared to the energy expansion. However, $K_{\infty}$ can be
reliably extracted in these specific cases. In realistic situations,
one also encounters modes of density vibrations, other than the
scaling and pure bulk modes, depending upon the mass region under
consideration.  When this dynamical effect is taken into account, the
nuclear compression modulus shows an `up-turn' behaviour below mass
number about 120, suggesting the inapplicability of the LDM expansion
of $K_A$ over this mass region.

\newpage
\medskip
\centerline {\bf APPENDIX }

Here an attempt is made to obtain a generalised empirical relation
between the nuclear compression modulus $K_A$ and the experimental
breathing-mode energies $E_{gmr}$. Until now, a well-defined relation
$E_{gmr} = \sqrt {{\hbar^2} K_A/ \ (m <r^2 >) } $,  is  available only
in the case of scaling model.

Brack and Stocker\cite{Br83} using a variational hydrodynamical
approach found that, the mass paramter $B(\beta_c, A)$
shows a regular behaviour with respect to both coupling paramter
$\beta_c$ and mass number $A$.
Using their results displayed in Table 1 of Ref.\cite{Br83}, we have
plotted $B(\beta_c, A)/ m<r^2> $ as a function of $A^{-1/3}$ in
Fig.(5).
It can be seen that the general mass paramter $B(\beta_c,A)$
expressed in terms of the scaling model value $m < r^2 >$ varies
quite smoothly with respect to A. Hence, $B(\beta_c,A)$ can be
expressed as
\begin{eqnarray}
B(\beta_c,A) &\sim & m<r^2> f(A)
\nonumber
\end{eqnarray}
where $f(A) = c_0 + c_1 A^{-1/3} + \cdots $ can be a polynomial in
$A^{-1/3}$.
The crucial function $f(A)$ can be determined from hydrodynamical
calculations, which will be reported elsewhere. Once $f(A)$ is known,
one can determine $K_A$ from GMR data using the relation
\begin{eqnarray}
h\omega &=& {\sqrt { K_A (\beta_c)\over B(\beta_c,A)} }
        \simeq {\sqrt { K_A (\beta_c)\over m<r^2> \cdot f(A)} } \nonumber
\end{eqnarray}
Thus, a generalised expression relating $K_A$ and $E_{gmr}$,
analogous to the well-known scaling relation, is proposed. This
proposition is on sound footing as the hydrodynamical studies are
successful in describing the GMR data.

\vfill
\newpage
\centerline {\bf TABLE CAPTIONS}
\vskip 1.0 true cm
\noindent {\bf Table I:}\
{Values of the various coefficients in the LDM
expansion of the energy(27) for the Skyrme forces SkM$^*$, SkA
and S3 using the local ETF functional with $\hbar^{2}$ terms
and a pure Fermi-function for the density distribution(3). All
quantities are in MeV. }
\vskip 1.0 true  cm
\noindent {\bf Table II:}\
{ Values of the various coefficients in the LDM
expansion of the compressibility(34) for the Skyrme forces SkM$^*$, SkA
and S3 using the local ETF functional with $\hbar^{2}$ terms
and a pure Fermi-function for the density distribution .
Three values of $\beta_c$ have been used.  All quantities are in MeV. }
\vskip 1.0 true cm
\noindent {\bf Table III:}\
{ Comparison of the exact values of the total
energy per nucleon $E/A$ (5) with the analytically determined values
using the LDM expansion of $E/A$, given by Eq.(27), for four
representative mass numbers $A$, and for the Skyrme forces SkM$^*$,
SkA and S3. All quantities are in MeV.}
\vskip 1.0 true cm
\noindent {\bf Table IV:}\
{ Comparison of the exact values of the finite
nuclear compression modulus $K_{A}$ (30) obtained using $\beta_c =0$
with the analytically determined values
using the LDM expansion of $K_{A}$ , given by Eq.(34), for four
representative mass numbers $A$, and for the Skyrme forces SkM$^*$, SkA and S3.
All quantities are in MeV.}
\vskip 1.0 true cm
\noindent {\bf Table V:}\
{ Values of the different terms contributing to
the total energy per nucleon obtained using the Skyrme forces
SkM$^*$, SkA and S3, for two representative mass numbers A. All
quantities are in MeV. }
\vskip 1.0 true cm
\noindent {\bf Table VI:}\
 {Values of the different terms contributing to
the nuclear incompressibility obtained using the Skyrme forces
SkM$^*$, SkA and S3 with $\beta_c =0$, for two representative mass
numbers A. All quantities are in MeV.}
\vskip 1.0 true cm
\noindent {\bf Table VII:}\
{ Values of the parameters obtained from a
least-squares fit to the exact values of energy per nucleon $E/A$
obtained for symmetric systems using the SkA force .
The number of data points used is 210, in the mass region $ 40
\le A \le 250 $ .
The first row gives the exact values for the various
coefficients obtained in our analytical model. All quantities
are in MeV.}
\vskip 1.0 true cm
\noindent {\bf Table VIII:}\
{ Values of the parameters obtained from a
least-squares fit to the exact values of nuclear
incompressibility $K_A$
obtained for symmetric systems using the SkA force with $\beta_c=0$.
The number of data points used is 210, in the mass region $ 40
\le A \le 250 $ .
The first row gives the exact values for the various
coefficients obtained in our analytical model. All quantities
are in MeV.}
\vskip 1.0 true cm
\noindent {\bf Table XI:}\
{Values of the parameters obtained from a
least-squares fit to the exact values of nuclear
incompressibility $K_A$ obtained for asymmetric systems
with Coulomb interaction, using the SkA force and taking
$\beta_c=0$.  The number of data points used is 210, in the mass
region $ 40 \le A \le 250 $ . The first row gives the exact
values for the various coefficients. Value of $K_{s\beta}$
obtained within the scaling model is taken from Ref.\cite{Na90}.
All quantities are in MeV.}
\vfill
\eject
\centerline {\bf FIGURE CAPTIONS}

\noindent {\bf FIG. 1}
 Values of the mass paramter $B(\beta_c, A)$
expressed in terms of $m<r^2>$, obtained in a hydrodynamical
calculation\cite{Br83} with SkM$^*$ force  is plotted versus $A^{-1/3}$,
where $m$ is the nucleon mass and $<r^2>$ is the root mean square
radius.

\noindent {\bf FIG. 2}
 Values of the ratio of the surface compressibility
coefficient $K_s$ to the nuclear matter incompressibility
$K_{v}$ obtained in our analytical model using
SkM$^*$ force is shown as a function of the coupling paramter $\beta_c$
{}.

\noindent {\bf FIG. 3}
Same as Fig. 2, but for the curvature
compressibility coefficient $K_c$.

\noindent {\bf FIG. 4}
Values of the ratio of finite nuclear compression
modulus $K_A$ to $K_v$ obtained including the dynamical effect\ ($A-$
dependence of $\beta_c$) is shown versus $A^{-1/3}$. The
force used is SkM$^*$.

\noindent {\bf FIG. 5}
Values of the ratio of finite nuclear compression
modulus $K_A$ to $K_v$ obtained for three particular values of
$\beta_c$ is shown versus $A^{-1/3}$. The force used is SkM$^*$.

\vfill
\newpage
\begin{table}
\centerline {\bf Table I}
\vskip 0.2 in
\begin{center}
\begin{tabular}{|c|c|c|c|}
\hline
\multicolumn{1}{|c|}{} &
\multicolumn{1}{|c|}{SkM$^*$} &
\multicolumn{1}{|c|}{SkA} &
\multicolumn{1}{|c|}{S3} \\
\hline
$a_v$&$-$15.79&$-$16.01&$-$15.87\\
$a_s$&19.08&19.95&18.90\\
$a_c$&10.24&9.67&6.87\\
$a_0$&$-$12.21&$-$11.42&$-$7.24\\
\hline
\end{tabular}
\end{center}
\end{table}
\vskip 0.1 true cm
\begin{table}
\centerline {\bf Table II}
\vspace{.3in}
\begin{center}
\begin{tabular}{|c|c|c|c|c|c|}
\hline
\multicolumn{1}{|c|}{Force} &
\multicolumn{1}{|c|}{$\beta_c$} &
\multicolumn{1}{|c|}{$K_s$(MeV)} &
\multicolumn{1}{|c|}{$K_c$(MeV)} &
\multicolumn{1}{|c|}{$K_o$(MeV)} &
\multicolumn{1}{|c|}{$ K_s/K_v $} \\
\hline
SkM$^*$&0&$-$406.1&$-$109.9&568.1&$-$1.87\\
($K_v$ = 217 MeV)&$-$1/3&$-$231.0&$-$129.3&138.1&$-$1.07\\
&$-$1/2&$-$129.1&$-$118.0&$-$69.1&$-$0.6\\
\hline
SkA&0&$-$484.6&$-$123.7&595.0&$-$1.84\\
($K_v$ = 263 MeV)&$-$1/3&$-$295.8&$-$145.7&167.8&$-$1.12\\
&$-$1/2&$-$186.4&$-$138.0&$-$43.8&$-$0.71\\
\hline
S3&0&$-$570.3&$-$114.1&452.5&$-$1.60\\
($K_v$ = 356 MeV)&$-$1/3&$-$389.6&$-$140.1&156.2&$-$1.10\\
&$-$1/2&$-$285.1&$-$142.7&5.26&$-$0.8\\
\hline
\end{tabular}
\end{center}
\end{table}
\vfill
\vfill
\eject
\begin{table}
\centerline {\bf Table III}
\vspace {.4in}
\begin{center}
\begin{tabular}{|c|c|c|c|}
\hline
\multicolumn{1}{|c|}{A} &
\multicolumn{1}{|c|}{SkM$^*$} &
\multicolumn{1}{|c|}{SkA} &
\multicolumn{1}{|c|}{S3} \\
\multicolumn{1}{|c|}{} &
\multicolumn{1}{|c|}{Exact \quad Eq.(27)} &
\multicolumn{1}{|c|}{Exact \quad Eq.(27)} &
\multicolumn{1}{|c|}{Exact \quad Eq.(27)} \\
\hline
40&  $-$9.72 \quad $-$9.64&  $-$9.70 \quad $-$9.64&$-$9.96 \quad $-$9.94 \\
100&$-$11.35 \quad $-$11.33&$-$11.39\quad $-$11.38&$-$11.56\quad$-$11.55\\
150&$-$11.93 \quad $-$11.92&$-$12.00\quad $-$12.00&$-$12.12\quad$-$12.12\\
200&$-$12.30 \quad $-$12.39&$-$12.38\quad $-$12.37&$-$12.47\quad$-$12.47\\
\hline
\end{tabular}
\end{center}
\end{table}
\vskip 2.0 true cm
\begin{table}
\centerline {\bf Table IV}
\vspace {.4in}
\begin{center}
\begin{tabular}{|c|c|c|c|}
\hline
\multicolumn{1}{|c|}{A} &
\multicolumn{1}{|c|}{SkM$^*$} &
\multicolumn{1}{|c|}{SkA} &
\multicolumn{1}{|c|}{S3} \\
\multicolumn{1}{|c|}{} &
\multicolumn{1}{|c|}{Exact \quad Eq.(34)} &
\multicolumn{1}{|c|}{Exact \quad Eq.(34)} &
\multicolumn{1}{|c|}{Exact \quad Eq.(34)} \\
\hline
40&103.4 \quad 102.7&126.9\quad 125.9&191.3\quad 190.3\\
100&130.2 \quad 129.7&159.5\quad 159.1&232.3\quad 231.9\\
150&140.4 \quad 140.1&172.0\quad 171.7&247.4\quad 247.1\\
200& 147.1 \quad 146.8 &180.0\quad 179.8& 257.1\quad 256.9 \\
\hline
\end{tabular}
\end{center}
\end{table}
\vfill
\eject
\begin{table}
\centerline {\bf Table V}
\vspace {.4in}
\begin{center}
\begin{tabular}{|c|c|c|c|c|c|}
\hline
\multicolumn{1}{|c|}{Force} &
\multicolumn{1}{|c|}{A} &
\multicolumn{1}{|c|}{$a_v$} &
\multicolumn{1}{|c|}{$a_sA^{-1/3}$} &
\multicolumn{1}{|c|}{$a_cA^{-2/3}$} &
\multicolumn{1}{|c|}{$a_0A^{-1}$} \\
\hline
\quad&40&$-$15.79&5.58&0.8755&$-$0.3052\\
SkM$^*$&200&$-$15.79&3.26&0.2994&$-$0.0611\\
\hline
\quad&40&$-$16.01&5.83&0.8268&$-$0.2855\\
SkA&200&$-$16.01&3.41&0.2828&$-$0.0571\\
\hline
\quad&40&$-$15.87&5.53&0.5874&$-$0.1810\\
S3&200&$-$15.87&3.23&0.2009&$-$0.0362\\
\hline
\end{tabular}
\end{center}
\end{table}
\vskip 1.0 true cm
\begin{table}
\centerline {\bf Table VI}
\vspace {.4in}
\begin{center}
\begin{tabular}{|c|c|c|c|c|c|}
\hline
\multicolumn{1}{|c|}{Force} &
\multicolumn{1}{|c|}{A} &
\multicolumn{1}{|c|}{$K_v$} &
\multicolumn{1}{|c|}{$K_sA^{-1/3}$} &
\multicolumn{1}{|c|}{$K_cA^{-2/3}$} &
\multicolumn{1}{|c|}{$K_0A^{-1}$} \\
\hline
\quad &40&216.6&$-$118.7&$-$9.40&14.20\\
SkM$^*$ &200&216.6&$-$69.4&$-$3.21&2.84\\
\hline
\quad &40&263.3&$-$141.7&$-$10.58&14.88\\
SkA&200&263.3&$-$82.9&$-$3.62&2.98\\
\hline
\quad &40&355.5&$-$166.8&$-$9.76&11.31\\
S3&200&355.5&$-$97.5&$-$3.34&2.26\\
\hline
\end{tabular}
\end{center}
\end{table}
\eject
\begin{table}
\centerline {\bf Table VII}
\vspace {.4in}
\begin{center}
\begin{tabular}{|c|c|c|c|c|}
\hline
\multicolumn{1}{|c|}{No. of } &
\multicolumn{1}{|c|}{$a_v$} &
\multicolumn{1}{|c|}{$a_s$} &
\multicolumn{1}{|c|}{$a_c$} &
\multicolumn{1}{|c|}{$a_0$} \\
\multicolumn{1}{|c|}{Para.} &
\multicolumn{1}{|c|}{} &
\multicolumn{1}{|c|}{} &
\multicolumn{1}{|c|}{} &
\multicolumn{1}{|c|}{} \\
\hline
\quad&$-$16.01&19.95&9.67&$-$11.42\\
2&$-$16.2$\pm$0.0003&22.1$\pm$0.002&\quad&\quad\\
3&$-$16.2$\pm$0.002&22.2$\pm$0.02&$-$0.30$\pm$0.04&\quad\\
4&$-$16.0$\pm$0.0001&19.7$\pm$0.001&11.6$\pm$0.007&$-$18.2 $\pm$ 0.01\\
\hline
\end{tabular}
\end{center}
\end{table}
\vskip 1.0 true cm
\begin{table}
\centerline {\bf Table VIII}
\vspace {.4in}
\begin{center}
\begin{tabular}{|c|c|c|c|c|}
\hline
\multicolumn{1}{|c|}{No. of } &
\multicolumn{1}{|c|}{$K_v$} &
\multicolumn{1}{|c|}{$K_s$} &
\multicolumn{1}{|c|}{$K_c$} &
\multicolumn{1}{|c|}{$K_0$} \\
\multicolumn{1}{|c|}{Para.} &
\multicolumn{1}{|c|}{} &
\multicolumn{1}{|c|}{} &
\multicolumn{1}{|c|}{} &
\multicolumn{1}{|c|}{} \\
\hline
\quad&263.3&$-$484.6&$-$125.8&604.4\\
2&256.4$\pm$0.14&$-$447.7$\pm$0.69&\quad&\quad\\
3&269.0$\pm$0.07&$-$569.1$\pm$0.65&282.1$\pm$1.5&\quad \\
4&263.7$\pm$0.24&$-$494.0$\pm$3.4&$-$67.1$\pm$15.8&531.2$\pm$24.0\\
\hline
\end{tabular}
\end{center}
\end{table}
\vfill
\eject
\begin{table}
\centerline {\bf Table IX}
\vspace {.4in}
\begin{center}
\begin{tabular}{|c|c|c|c|c|c|}
\hline
\multicolumn{1}{|c|}{$K_v$} &
\multicolumn{1}{|c|}{$K_s$} &
\multicolumn{1}{|c|}{$K_{\beta}$} &
\multicolumn{1}{|c|}{$K_{Coul}$} &
\multicolumn{1}{|c|}{$K_c$} &
\multicolumn{1}{|c|}{$K_{s\beta}$}\\
\multicolumn{1}{|c|}{} &
\multicolumn{1}{|c|}{} &
\multicolumn{1}{|c|}{} &
\multicolumn{1}{|c|}{} &
\multicolumn{1}{|c|}{} &
\multicolumn{1}{|c|}{} \\
\hline
263.3&$-$484.6&$-$441.1&$-$5.14&$-$125.8& $\sim$ 875 \\
252.3$\pm$2.0&$-$428.4$\pm$5.6&$-$240.8$\pm$7.0&$-$4.7$\pm$0.19&\quad&\quad\\
237.5$\pm$5.8&$-$391.2$\pm$14.7&$-$394.6$\pm$56.8&$-$3.28$\pm$0.55&\quad &
1009.8$\pm$370.0\\
368.7$\pm$17.3&$-$1108.4$\pm$100.5&$-$423.3$\pm$28.0&$-$9.4$\pm$0.71&
1126.5$\pm$166.3&\\
\hline
\end{tabular}
\end{center}
\end{table}
\vfill
\end{document}